\documentclass[prb,aps,twocolumn,floatfix]{revtex4}
\usepackage{graphicx}% Include figure files
  \usepackage{amsmath}
\usepackage{amssymb}
\usepackage{amsfonts}
\usepackage{bm}
\begin{document}

\newcommand{\beq}{\begin{eqnarray}}
\newcommand{\eeq}{\end{eqnarray}}
\newcommand{\bk}{{\bf k}}

\date{\today}
\title{Superfluid and supersolid phases of lattice bosons with ring-exchange interaction}
\author{Robert Schaffer}
\affiliation{Department of Physics and Astronomy, University of Waterloo, Ontario, N2L 3G1, Canada} 
\author{Anton A. Burkov}
\affiliation{Department of Physics and Astronomy, University of Waterloo, Ontario, N2L 3G1, Canada} 
\author{Roger G. Melko}
\affiliation{Department of Physics and Astronomy, University of Waterloo, Ontario, N2L 3G1, Canada}

\begin{abstract}
We examine the superfluid phase of a hard-core boson model with nearest-neighbor exchange 
$J$ and four-particle ring-exchange $K$ at half-filling on the square lattice.   At zero temperature we find that the superfluid in the pure-$J$ model is quickly destroyed by the inclusion of  negative-$K$ ring-exchange interactions, favoring a state with a $(\pi,\pi)$ ordering wavevector.  Minimization of the mean-field energy suggests that a supersolid state with coexisting superfluidity, charge-density wave, and valence-bond-like order is formed.   
We also study the behavior of the finite-$T$ Kosterlitz-Thouless phase transition in the superfluid phase, by forcing the Nelson-Kosterlitz universal jump condition on the finite-$T$ spin wave superfluid density.  Away from the pure $J$ point, $T_{KT}$ decreases rapidly for negative 
$K$, while for positive $K$, $T_{KT}$ reaches a maximum at some $K \neq 0$ in agreement with recent quantum Monte Carlo simulations.
\end{abstract}
\maketitle

%\newpage

\section{Introduction}

The road to discovering new quantum phenomena such as exotic phases or unconventional quantum phase transitions is paved by the understanding of quantum Hamiltonians with competing microscopic interactions.  
Recent work has shown that competing kinetic energy terms, particularly four-particle ``ring-exchange'', is a fertile venue for the study of valence-bond solids and unconventional quantum phase transitions.\cite{Arun1,JK_1st,Arun2,rick1}

Long relegated to toy models of easy-plane spin systems or simple boson theories, the understanding of U(1) Hamiltonians has resurfaced as critical for the engineering of exotic quantum phases of ultracold atoms in optical lattices.   Buchler {\it et al.}\cite{Buchler} presented the design of a ring-exchange interaction for bosonic cold atoms in two-dimensional (2D) square optical lattices.  There, a conventional nearest-neighbor hopping term competes with a four-particle ring-exchange, which arises out of the hopping of boson pairs on the corners of square lattice plaquettes to an intermediate molecular state and back to the (opposite) plaquette corners.  The Hamiltonian proposed in Ref.~\onlinecite{Buchler} can be written,
\begin{equation}
H = - J\sum_{\langle ij \rangle} \left({ b^{\dagger}_i b^{}_j + b^{}_i b^{\dagger}_j  }\right) - K\sum_{\langle ijkl \rangle} \left({ b^{\dagger}_i b^{}_j b^{\dagger}_k b^{}_l + b^{}_i b^{\dagger}_j b^{}_k b^{\dagger}_l }\right), \label{JKham}
\end{equation}
where the first term is the usual boson hopping, and the second is the ring-exchange interaction.  Here, $i$ and $j$ are neighboring sites, lying opposite to sites $l$ and $k$ (respectively), which together form the basic square plaquette of the 2D lattice.
This model has drawn significant attention recently from proposals that it may harbor an exotic quantum liquid phase   that possesses d-wave correlations --  a so-called d-wave Bose liquid (DBL) phase\cite{DBL} -- including extensive numerical investigation using  exact diagonalization and the density  matrix renormalization group.\cite{Donna}

The main difficulty with the Hamiltonian Eq.~(\ref{JKham}) that prevents an exact solution in 2D for large systems (or the thermodynamic limit) is the presence of the prohibitive ``sign-problem'' in quantum Monte Carlo (QMC) simulations for the parameter regime $K<0$.  In fact, the same model in the parameter regime without the sign problem, $K>0$, has been solved by QMC previously both at half-filling\cite{JK_1st} and in the presence of a symmetry-breaking chemical potential.\cite{JK_h}   In this model, it was demonstrated\cite{JK_1st} that the superfluid phase that dominates for large $J$ is destroyed by increasing the magnitude of the ring-exchange interaction, which realizes first a $(\pi,0)$ valence-bond solid (VBS) for $8<K/J<14$, and a  $(\pi,\pi)$ charge-density-wave (CDW) for $K>14$.
  The intermediate superfluid to VBS quantum phase transition was studied intensely\cite{JKannl}  as one of the first candidates for a {\it deconfined} quantum critical point.\cite{DQCP1}  However, other interesting behavior occurs in the model, in particular in the experimentally relevant regime of finite-temperatures, where relatively little attention have been paid.\cite{JKannl}   

In order to elucidate the mechanism by which the superfluid phase, which occurs for dominant $J$ in Eq.~(\ref{JKham}),  is destroyed by the four-site $K$ interaction, we explore in this paper the behavior of the model using linear spin-wave (SW) theory.  First, we calculate the dispersion and superfluid density $\rho_s$ at $T=0$, where its behavior as a function of $K$ indicates the realization of a phase with $(\pi,\pi)$ symmetry for $K/J < -2$.  Exploring the ordering nature of this phase in mean-field theory suggests a $(\pi,\pi)$ modulated in-plane order parameter, the mean-field indicator of a supersolid state, where superfluidity is expected to coexist with both CDW and also valence-bond-like order.
In the superfluid phase, we also  calculate the superfluid density at finite-temperature, and estimate the position of the Kosterlitz-Thouless (KT) transition using the universal jump condition.  Interestingly, for $K>0$ the linear spin-wave theory reproduces the QMC result\cite{JKannl} that the maximum temperature of the KT transition does {\it not} occur when $K=0$, rather it happens at some intermediate value of $K/J$.  

\section{Linear spin wave theory at $T=0$}
In order to study the destruction of the superfluid phase by the competing ring-exchange interaction parameterized by $K$, we begin by examining the behavior of the system at $T=0$ in a linear spin-wave analysis.  To begin, it is helpful to write our hard-core boson Hamiltonian as the analogous two-dimensional spin-1/2 XY model using the standard mapping: $b^{\dagger}_{i} \rightarrow S^{+}_i$ and $b_{i} \rightarrow S^{-}_i$. This transforms the Hamiltonian Eq.~(\ref{JKham}) into 
\begin{eqnarray}
\lefteqn{H = -J\sum_{\langle ij \rangle} (S^{+}_{i}S^{-}_j+S^{-}_iS^{+}_j)} \nonumber\\
& & - K\sum_{\langle ijkl \rangle} (S^{+}_{i}S^{-}_jS^+_kS^-_l+S^{-}_iS^{+}_jS^-_kS^+_l), \label{JKxy}
\end{eqnarray}
which can be written with $S^x$ and $S^y$ operators as
\begin{eqnarray}
H =&-&2J\sum_{\langle ij \rangle} (S^{x}_{i}S^{x}_j+S^{y}_iS^{y}_j) \label{xx} \\
 &-& 2K\sum_{\langle ijkl \rangle} \Big(S^{x}_{i}S^{x}_jS^x_kS^x_l+S^{x}_iS^{x}_jS^y_kS^y_l \nonumber \\
 &-& S^{x}_{i}S^{y}_jS^x_kS^y_l+S^{x}_iS^{y}_jS^y_kS^x_l +S^{y}_{i}S^{x}_jS^x_kS^y_l  \nonumber \\
 &-& S^{y}_iS^{x}_jS^y_kS^x_l + S^{y}_{i}S^{y}_jS^x_kS^x_l+S^{y}_iS^{y}_jS^y_kS^y_l\Big). \nonumber
\end{eqnarray}
As mentioned above, the groundstate of the pure-$J$ Hamiltonian is a bosonic superfluid, or an in-plane ferromagnet in the spin language, with an order parameter $\langle S^x \rangle \neq 0$ at zero temperature.  We therefore perform our spin-wave expansion around this ordered state, treating the $K$ term of the Hamiltonian as a perturbation.  As first demonstrated by Gomez-Santos and Joannopoulos,\cite{GomezJ} the proper Holstein-Primakoff representation in the case of the XY model is
\begin{eqnarray}
S^x_i &\approx& \frac{1}{2} - a_i^\dagger a_i,  \nonumber\\
S^y_i &\approx& \frac{1}{2i}(a_i^\dagger - a_i),
\end{eqnarray}
giving us a leading order approximation to the Hamiltonian in terms of bosonic spin-wave operators $a_i$ and $a^{\dagger}_i$. 
Writing the Hamiltonian in Fourier space gives the linear spin-wave Hamiltonian
\begin{eqnarray}
H = H_{MF} + \sum_{\bk} \Big[ &A_{\bk}& ( a_{\bk}^\dagger a_{\bk} + a_{-{\bk}}^\dagger a_{-{\bk}}) \nonumber\\
+ &B_{\bk}&(a_{\bk}^\dagger a_{-{\bk}}^\dagger + a_{\bk} a_{-{\bk}})\Big] . \label{SW_1}
\end{eqnarray}
Here, the mean-field energy is
\beq
H_{MF} = -JN - \frac{KN}{8}
\eeq
while the coefficients $A_{\bk}$ and $B_{\bk}$
\begin{eqnarray}
A_{\bk} &=& JU_{\bk} + KV_{\bk}\\
B_{\bk} &=& JW_{\bk} + KX_{\bk}
\end{eqnarray}
are defined in terms of 
\begin{eqnarray}
U_{\bk} &=& 2 - \frac{1}{2}\gamma_{\bk} \label{Uu} \\
V_{\bk} &=& \frac{1}{2} - \frac{1}{4}\gamma_{\bk} +\frac{1}{4}\cos k_x \cos k_y \label{Vv} \\
W_{\bk} &=& \frac{1}{2}\gamma_{\bk} \label{Ww} \\
X_{\bk} &=& \frac{1}{4}\gamma_{\bk} - \frac{1}{4}\cos k_x \cos k_y \label{Xx}
\end{eqnarray}
where
\beq
\gamma_{\bk} = \cos k_x +\cos k_y.
\eeq
This form for the SW Hamiltonian reduces to that obtained by Bernardet {\it et.~al.~}\cite{TroyerSWT} in the limit of the simple XY model ($K=0$) and the absence of an external magnetic field.  Following Ref.~\onlinecite{TroyerSWT}, we diagonalize Eq.~(\ref{SW_1}) using a Bogoliubov transformation,
\beq
a_{\bk} = u_{\bk}\alpha_{\bk} - v_{\bk}\alpha^\dagger_{-{\bk}}, \quad a_{\bk}^\dagger = u_{\bk}\alpha_{\bk}^\dagger - v_{\bk}\alpha_{-{\bk}}.
\eeq
where $\alpha_{\bk}$ and $\alpha_{\bk}^{\dagger}$ are destruction and creation operators for quasiparticles of momentum ${\bk}$.  The bosonic commutation relations are satisfied if
\beq
u_\bk = \cosh(\varphi_\bk), \hspace{0.5cm} v_\bk = \sinh (\varphi_\bk)
\eeq
where $\varphi_k$ is determined by the requirement that it diagonalizes our Hamiltonian.  This yields
\begin{eqnarray}
u_\bk^2 &=& \frac{1}{2} \left({  \frac{A_\bk}{\sqrt{A_\bk^2-B_\bk^2}} + 1 }\right), \\
v_\bk^2 &=& \frac{1}{2} \left({  \frac{A_\bk}{\sqrt{A_\bk^2-B_\bk^2}} - 1 }\right),
\end{eqnarray}
and thus the diagonalized spin-wave Hamiltonian\cite{TroyerSWT}
\begin{eqnarray}
H = &H_{MF}& + \sum_{\bk}(\sqrt{A_{\bk}^2 - B_{\bk}^2} - A_{\bk}) \label{DiagSWT} \\
&+& \sum_{\bk}\sqrt{A_{\bk}^2 - B_{\bk}^2}(\alpha^\dagger_{\bk}\alpha_{\bk} + \alpha_{-\bk}^\dagger\alpha_{-\bk}). \nonumber
\end{eqnarray}
It is the diagonal SW Hamiltonian, in this generalized form, that we now proceed to use to analyze the behavior of our superfluid phase at zero temperature.

\subsection{Dispersion}

\begin{figure}
{
\includegraphics[width=2.8in]{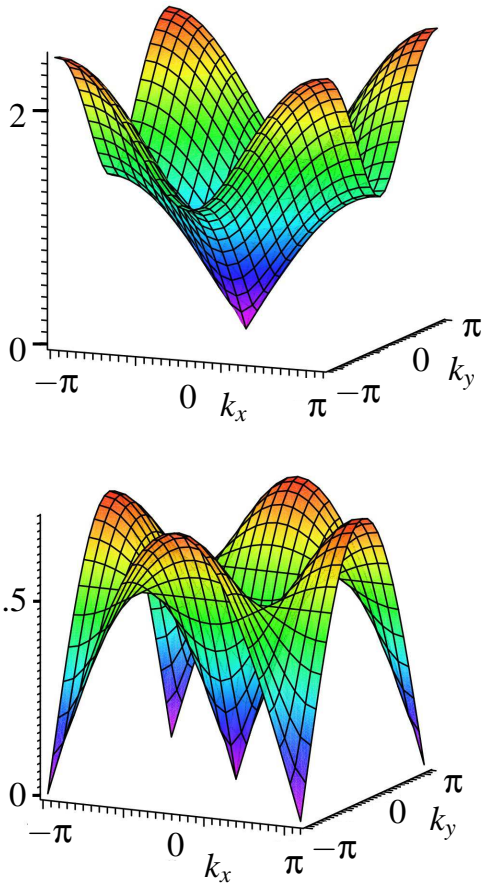}
\caption{(Color online)  The dispersion $\omega_{\bf k}$, as a function of ${\bf k}=(k_x,k_y)$,  for $K/J = 2$ (top) and $K/J = -2$ (bottom).
\label{disp}}
}
 \end{figure}

We begin by studying the dispersion, which is obtained immediately from Eq.~(\ref{DiagSWT})
\beq
\omega_{\bk} = 2\sqrt{A_{\bk}^2 - B_{\bk}^2}.
\eeq
This can be examined as a function of $K/J$ for the Hamiltonian Eq.~(\ref{JKxy}).  For $K=0$ and ${\bf k}\rightarrow 0$, one reproduces the expected linear dispersion,\cite{TroyerSWT} a feature which survives for moderate $K>0$ (at least to first-order in the SW theory) --  see Fig.~\ref{disp}.   No soft modes develop in the $K>0$ dispersion until very large values of $K/J \approx 10^3$, which is well above the critical value of $K/J=7.91$ where a phase transition to a $(\pi,0)$ VBS is known from QMC.\cite{JK_1st}
In contrast is the behavior of the dispersion in the $K < 0$ regime, which is intractable to QMC methods due to the negative sign problem. In this case, the dispersion reveals the development of soft modes at $\bk=(\pi,\pi)$.  The value of  $\omega_{(\pi,\pi)}$ tends towards 0 as $K/J$ approaches $-2$, as illustrated in Fig.~\ref{disp}.   This indicates that an ordered phase with $(\pi, \pi)$ symmetry is realized for sufficiently large $| K |$.  The nature of this ordering is discussed in Section \ref{SSsection}.

\subsection{The zero-temperature superfluid density}

The most important quantity characterizing the superfluid phase is the superfluid density (or spin stiffness), defined as the second derivative of the free energy with respect to a uniform twist $\phi$ of the in-plane spin components across the system. 
At zero-temperature, one replaces the free energy by the groundstate energy, hence we use the definition 
\beq
\rho_s = \frac{\partial^2E(\phi)}{\partial^2\phi},
\eeq
which is valid for $\phi \rightarrow 0$. This is equivalent to finding the incremental energy per spin resulting from the twist,
\beq
\frac{\Delta E}{N} = \frac{\langle H(\phi)\rangle}{N} - \frac{\langle H(0)\rangle}{N} = \frac{1}{2} \rho_{s}\phi^2.
\eeq
Thus, we are faced with the task of finding $H(\phi)$. To do this, we consider the spins in a site-dependent rotated reference frame, defined via the standard rotation operator about the $z$ axis.\cite{AWS_heisenberg}
The relevant transformations become
\begin{eqnarray}
S_j^x \quad & \rightarrow & \quad S^{x}_j\cos{\phi_j} - S^{y}_j\sin{\phi_j}, \nonumber\\
S_j^y \quad & \rightarrow & \quad S^{x}_j\sin{\phi_j} + S^{y}_j\cos{\phi_j},\label{invrot}
\end{eqnarray}
which can also be written in terms of the spin raising and lowering operators,
\begin{eqnarray}
S_j^+ \quad & \rightarrow & \quad S^{+}_je^{i\phi_j},\nonumber\\
S_j^- \quad & \rightarrow & \quad S^{-}_je^{-i\phi_j}.
\end{eqnarray}
We assume a uniform twist across the spins in the system, so that the twist $\phi_i - \phi_j \equiv \phi$,  where $i$ and $j$ are nearest neighbors in the $x$ or $y$ direction.  We find that, using the labeling of Eq.~(\ref{JKham}), $\phi_j = \phi_k, \phi_i = \phi_l$, and thus the $\phi$ dependence of the ring-exchange term cancels.  This gives for our twisted Hamiltonian
\begin{eqnarray}
H(\phi) = -2J\sum_{\langle ij \rangle}\big[(S^{x}_{i}S^{x}_j+S^{y}_iS^{y}_j)\cos{\phi}\nonumber\\
+ (S_i^xS_j^y - S^y_iS^x_j)\sin{\phi}\big],\nonumber\\
-K\sum_{\langle ijkl \rangle} (S^{+}_{i}S^{-}_jS^+_kS^-_l+S^{-}_iS^{+}_jS^-_kS^+_l).
\end{eqnarray}
The linear components of the term proportional to $\sin{\phi}$ cancel, so in linear spin wave theory the twist effectively scales the nearest neighbor exchange value $J \rightarrow J\cos{\phi}$.  Thus, $H(\phi)$ is obtained directly from the Hamiltonian Eq.~(\ref{SW_1}), with the redefinition of
\begin{eqnarray}
H_{MF}(\phi) &=& -JN\cos{\phi} - \frac{KN}{8}, \\
A_{\bk}(\phi) &=& JU_{\bk}\cos{\phi} + KV_{\bk},\\
B_{\bk}(\phi) &=& JW_{\bk}\cos{\phi} + KX_{\bk},
\end{eqnarray}
but with the coefficients $U_{\bk}$, $V_{\bk}$, $W_{\bk}$, and $X_{\bk}$ unchanged from Eqs.~({\ref{Uu}-\ref{Xx}}). Expanding $\cos{\phi}$ as $1-\phi^2/2$, and noting that at zero temperature, the thermal expectation value is
\beq
n_{\bk} = \langle \alpha_{\bk}^\dagger\alpha_{\bk}\rangle = \langle \alpha_{-\bk}^\dagger\alpha_{-\bk}\rangle = 0,
\eeq
we find that 
\begin{eqnarray}
\rho_s &=& \frac{J}{2} + \frac{1}{2N}\sum_{\bk}\Big[JU_{\bk} - (A_{\bk}^2 - B_{\bk}^2)^{-1/2} \label{rho0} \\
&\cdot&\big(J^2(U_{\bk}^2 - W_{\bk}^2) + JK(U_{\bk}V_{\bk} - W_{\bk}X_{\bk})\big)\Big] . \nonumber
\end{eqnarray}
Note that we have divided by a factor of two to account for the fact that Eq.~(\ref{DiagSWT}) is defined with respect to a twist in a single lattice bond, while Eq.~(\ref{invrot}) considers a twist in each bond in the lattice, a total of 2N bonds. 

\begin{figure}
{
\includegraphics[width=3.2in]{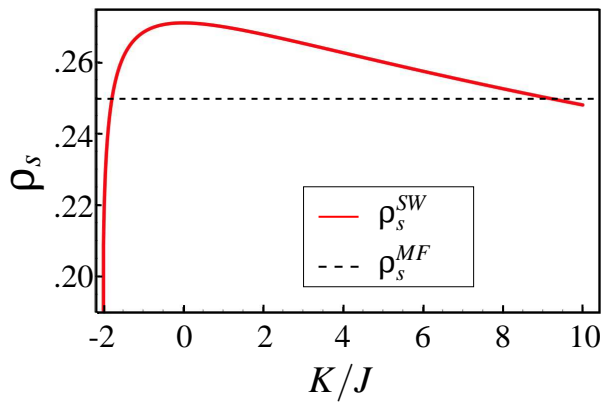}
\caption{(Color online) The superfluid density as a function of $K/J$.  The dashed line is the mean-field result, $\rho_s=0.25$.   The linear SW result for $K/J=0$ is $\rho_s=0.2709$, which can be compared to the best known exact results $\rho_s$ = 0.2696(2).\cite{AWShammer}
\label{ps}}}
 \end{figure}
 
The $T=0$ superfluid density calculated in this linear SW theory, Eq.~(\ref{rho0}), is plotted in Fig.~\ref{ps} for a range of $K/J$.   Note that, as in the above section, we have set $J=1/2$ to correspond to the usual definition of the XY spin model (when $K=0$).   The superfluid density curve has its maximum at $K/J=0$, with a value of $\rho_s = 0.2709$, which can be compared to the best numerical estimate for the groundstate spin stiffness in the XY model using finite-size scaling quantum Monte Carlo techniques, $\rho_s$ = 0.2696(2).\cite{AWShammer}  Away from this maximum at $K=0$, the superfluid density declines monotonically with increasing $|K|$.  On the positive $K$ side, $\rho_s$ decreases relatively gradually, only becoming zero for an extremely large value of $K \approx 10^{3}$.  This is consistent with the above results for the dispersion, which indicate that no soft modes develop for moderate values of positive $K$.  Again, exact QMC results have revealed a $T=0$ quantum phase transition to a valence-bond solid phase at $K/J \approx 7.91$,\cite{JK_1st} reiterating the limitations of the linear SW theory calculation in this regime.

On the negative-$K$ side, where no QMC results are available, the value of $\rho_s$ drops rapidly as it approaches -2.  In SW theory, there is a divergent negative contribution to $\rho_s$ at this value of $K/J=-2$.  This is of course an artifact of linear SW theory.   As explained in the next section, one actually expects $\rho_s$ to remain finite on the other side of the transition, where an ordered phase with $(\pi,\pi)$ symmetry is realized.

\begin{figure}
{
\includegraphics[width=1.8in]{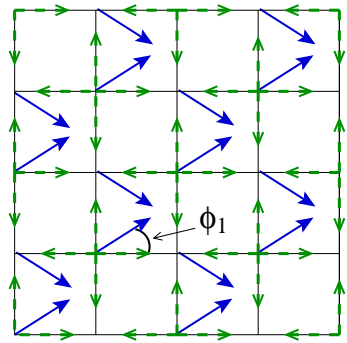}
\caption{(Color online)  The $(\pi,\pi)$ supersolid phase as dictated by mean-field theory, for $K/J<-2$.  Solid (blue) arrows show the angles of deviation of the classical spin vectors from the $x$ direction (the out-of plane deviation is zero).  Dashed (green) arrows are the corresponding bosonic charge currents $\mathcal{J}$, Eq~(\ref{Jcurr}).
\label{SSfig}}
}
 \end{figure}

\section{Supersolid phase for $K<0$} \label{SSsection}

Examining our linear SW theory results of the previous section, for $K/J<-2$ we expect the development of a phase with $(\pi,\pi)$ as the ordering wavevector.  We can further elucidate the nature of the ordering in this phase by minimizing the mean-field energy of the ring-exchange Hamiltonian, Eq.~(\ref{JKham}), under the constraint of order with $(\pi,\pi)$ symmetry.
By representing the quantum spins as classical spin vectors, 
\beq
S_i^x &=& \frac{1}{2} \sin \theta_i \cos \phi_i, \\
S_i^y &=& \frac{1}{2} \sin \theta_i \sin \phi_i,  \\
S_i^z &=& \frac{1}{2} \cos \theta_i, 
\eeq
we rewrite Eq.~(\ref{xx}), which simplifies to the following Hamiltonian:
\begin{eqnarray}
H&=&-\frac{J}{2}\sum_{\langle ij \rangle} \sin \theta_i \sin \theta_j \cos ( \phi_i-\phi_j) \label{Hmft} \\
&-& \frac{K}{8} \sum_{\langle ijkl \rangle} \sin \theta_i \sin \theta_j \sin \theta_k \sin \theta_l \cos (\phi_i-\phi_j+\phi_k-\phi_l). \nonumber
\end{eqnarray}
Note that the angle of the classical vector from the $x$ axis, $\phi_i$, should not be confused with the phase twist variable defined in the previous section.
Restricting ourselves to half-filling, we look for solutions to the angles $\theta_i$ and $\phi_i$ in the form
\begin{eqnarray}
\theta_i &=& \theta_0 + \theta_1 e^{i {\boldsymbol \pi} \cdot {\bf r}_i }, \\
\phi_i &=& \phi_0 + \phi_1 e^{i {\boldsymbol \pi} \cdot {\bf r}_i },
\end{eqnarray}
(where $ {\boldsymbol \pi} = (\pi,\pi)$) by minimizing the classical energy Eq~(\ref{Hmft}).  The solution is
\beq
\theta_0 &=& \theta_1 = \phi_0 =  0 \\
\phi_1 &=& \frac{1}{2} \arccos \left({ \frac{2J}{K} }\right).
\eeq
This solution corresponds to a staggered canting of spins away from the $x$ axis, as illustrated in Fig.~\ref{SSfig}.  In the boson language, we can map this to a charge current using the definition
\beq
\mathcal{J}_{ij} = i \frac{J}{2} (S_i^+ S_j^- - S_i^- S_j^+) = J \sin(\phi_i - \phi_j). \label{Jcurr}
\eeq
This state would correspond to a staggered charge current as shown in Fig.~\ref{SSfig}.

In MFT, the above state does not contain any charge density modulation, since the out-of-plane angle $\theta_1$ is found to be zero.  However, one expects (e.g.~based on Landau-Ginzburg type symmetry arguments) such charge density modulation to appear once fluctuations are included.   Such a state is therefore a bosonic {\it supersolid} phase, with coexistent superfluidity and $(\pi,\pi)$ CDW order. 
Interestingly, from the modulation of the boson current operator $\mathcal{J}$ illustrated in Fig.~\ref{SSfig}, it is clear that the supersolid phase in this case has the curious additional feature of valence-bond-like order coexisting with CDW density modulation, a situation which does not typically occur in lattice supersolids arising from such simple Hamiltonians.  
It is also worth noting that the mechanism for formation of this supersolid is distinct from that of other hard-core boson supersolid phases at commensurate filling discussed previously,\cite{ss1,ss2,ss3,ss4} which are induced by the geometrical frustration of the lattice.

\section{Finite-temperature Kosterlitz-Thouless transition}

We turn now to a discussion of the superfluid phase in the J-K model at finite temperatures.  The goal of this section will be to map out the finite-temperature phase boundaries of the superfluid phase in our linear SW theory, to compare (at least in part) to the results obtained in other studies.  In the QMC work of Ref.~\onlinecite{JKannl}, the phase boundary reported has a non-trivial shape, reaching a maximum for a positive value of $K \neq 0$.  We are interested in whether this feature can be captured by our SW theory, therefore elucidating the physical mechanism by which this maximum at $K \neq 0$ occurs.  To do this, we first develop an expression for the superfluid density for $T>0$ (Section \ref{IVA}), then use it to estimate the SW Kosterlitz-Thouless transition through the universal jump condition (Section \ref{IVB}).

\subsection{The finite-temperature superfluid density} \label{IVA}

\begin{figure}
{
\includegraphics[width=3.2in]{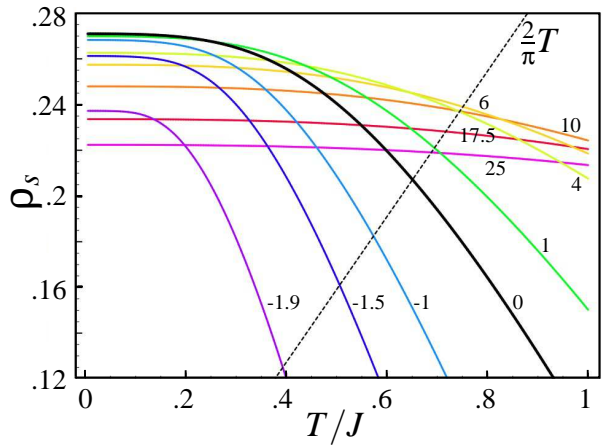}
\caption{(Color online) The superfluid density, calculated from Eq.~(\ref{TrhoF}), as a function of the temperature.  Lines are labeled by the parameter value $K/J$.   The universal jump (Eq.~(\ref{tkt})) is illustrated as a dashed line.
The exact value for the Kosterlitz-Thouless transition for $K/J=0$ is $T_{KT}=0.6860(2)$, from unbiased QMC calculations.\cite{RogXY}
\label{psvsT}}}
 \end{figure}

At finite temperatures, we may calculate the superfluid density by recalling its original definition as the response of the free energy with respect to a twist.  Calculating the partition function from Eq.~(\ref{DiagSWT}) we obtain the free energy
\begin{eqnarray}
F = H_{MF}(\phi) &+& \sum_{\bk}\left({\sqrt{A_{\bk}(\phi)^2 - B_{\bk}(\phi)^2} - A_{\bk}(\phi) }\right)\nonumber\\
&+& T \sum_{\bk}\ln{\left(1 - e^{-\omega_{\bk}(\phi)/T}\right)},
\end{eqnarray}
with the dispersion redefined as
\beq
\omega_{\bk}(\phi) = 2\sqrt{A_{\bk}(\phi)^2 - B_{\bk}(\phi)^2}.
\eeq
As noted previously, the  twist-dependent terms in $H(\phi)$ are proportional to $\cos{\phi}$, so we can write for $\phi \rightarrow 0$
\beq
F = F(\phi=0) + \frac{2N}{2}\rho_s\phi^2 + \ldots
\eeq
which is properly normalized with $2N$ being the number of lattice bonds.  Therefore, we obtain for the superfluid density
\begin{align} 
\rho_s(T) = &\lim_{\phi \rightarrow 0} \frac{1}{N}\frac{\partial F}{\partial\phi^2}\\
= &\lim_{\phi \rightarrow 0} \frac{J}{2} + \frac{1}{N}\sum_{\bk}\big(\frac{1}{2}\frac{\partial\omega_{\bk}(\phi)}{\partial\phi^2} - \frac{\partial A_{\bk}(\phi)}{\partial\phi^2}\big)\nonumber\\
&+ \frac{T}{N}\sum_{\bk}\frac{e^{-\omega_{\bk}/T}}{1- e^{-\omega_{\bk}/T}}\frac{1}{T}\frac{\partial\omega_{\bk}(\phi)}{\partial\phi^2},\\\nonumber
\end{align}
where the important limits are evaluated as follows:
%\begin{align}
\begin{eqnarray}
\lim_{\phi \rightarrow 0}  \frac{\partial A_{\bk}(\phi)}{\partial\phi^2} &=& -J\frac{U_{\bk}}{2}\\
\lim_{\phi \rightarrow 0} \frac{\partial B_{\bk}(\phi)}{\partial\phi^2} &=& -J\frac{W_{\bk}}{2}\\
\lim_{\phi \rightarrow 0} \frac{\partial \omega_{\bk}(\phi)}{\partial\phi^2} &=& \lim_{\phi \rightarrow 0} \frac{4}{\omega_{\bk}}\left({ A_{\bk}\frac{\partial A_{\bk}(\phi)}{\partial\phi^2} - B_{\bk}\frac{\partial B_{\bk}(\phi)}{\partial\phi^2} }\right)\nonumber\\
&=& -2J\frac{A_{\bk}}{\omega_{\bk}}U_{\bk} + 2J\frac{B_{\bk}}{\omega_{\bk}}W_{\bk} .
\end{eqnarray}
%\end{align}
Thus, this calculation yields the following expression for the superfluid density at finite-$T$:
\begin{align}
\rho_s(T) &= \frac{J}{2} + \frac{J}{2N}\sum_{\bk}U_{\bk} \label{TrhoF} \\
& - \frac{J}{N}\sum_{\bk}\left({ 1 + 2\frac{1}{e^{-\omega_{\bk}/T} -1}}\right)\left({\frac{A_{\bk}}{\omega_{\bk}}U_{\bk} - \frac{B_{\bk}}{\omega_{\bk}}W_{\bk}}\right).  \nonumber
\end{align}
This SW expression for $\rho_s$ as a function of temperature is plotted for several parameters $K/J$ in Fig.~\ref{psvsT}.

\subsection{Estimation of the Kosterlitz-Thouless transition} \label{IVB}

The superfluid density of Eq.~(\ref{TrhoF}), plotted as a function of $T$, decays slowly (see Fig.~\ref{psvsT}), crossing zero for relatively large temperatures.  However, as is well known, the XY model realizes a Kosterlitz-Thouless\cite{KT} (KT) transition in any two-dimensional lattice, which is manifest as a discontinuity in $\rho_s$ precisely at  $T_{KT}$.  This, of course, can not be captured by a simple spin wave theory.  
We can, however, take advantage of the fact that the expected discontinuity in $\rho_s$ in 2D models such as this obeys the so-called {\it universal jump} condition,
\beq
T_{KT} = \frac{\pi}{2} \rho_s(T_{KT}) \label{tkt},
\eeq
first found by Nelson and Kosterlitz.\cite{Ujump}  
One can thus hypothesize a reasonable estimate of $T_{KT}$ by solving $T = \pi \rho_s(T)/2$, using $\rho_s(T)$ from our 
spin-wave theory.  An important test of this idea can be performed at $K=0$ and $J=1/2$, the parameter values for the XY model.  This is illustrated in Fig.~\ref{psvsT}, where the crossing point of the dashed line and the curve for $K/J=0$ is our SW theory solution to Eq.~(\ref{tkt}).  The value of $T_{KT}$ obtained by this procedure is 0.6507, remarkably close to the exact result of $T_{KT}=0.6860(2)$, obtained from unbiased QMC calculations.\cite{RogXY}

\begin{figure}
{
\includegraphics[width=3.2in]{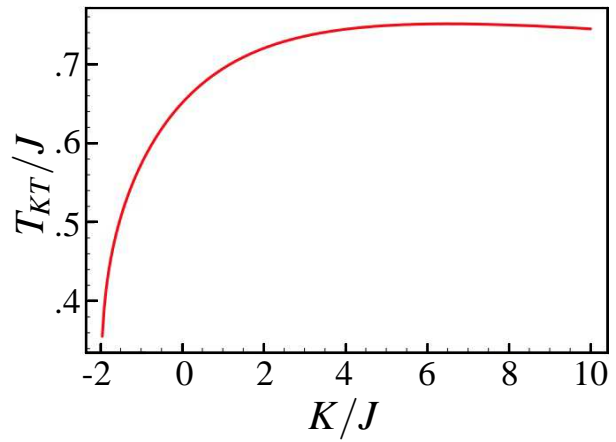}
\caption{(Color online) The KT transition phase boundary as calculated in linear SW theory.  A maximum of $T_{KT}/J = 0.751$ occurs at approximately $K/J=6.5$.   
\label{tktvsk}}}
 \end{figure}

We thus generalize this procedure of calculating $T_{KT}$ in our linear SW theory to nonzero values of $K/J$.  Results are plotted in Fig.~\ref{tktvsk}.  As is evident by studying Fig.~\ref{psvsT} closely, the SW theory reproduces the remarkable trend that $T_{KT}$ {\it increases} for small to moderate values of $K>0$.  In Fig.~\ref{psvsT}, the maximum in $T_{KT}$ occurs at $K/J\approx 6.5$, before beginning to drop slowly.  The phase boundary for the KT transition drops to zero for the large value of $K/J\approx 10^3$, well outside of the expected range of validity for the SW calculation.  Clearly, a simple spin-wave theory cannot capture the physics of the superfluid-VBS quantum phase transition that is observed in QMC simulations.

On the negative-$K$ side, the phase boundary drops rapidly to zero as $K/J$ approaches $-2$, confirming the expectation from $T=0$ SW theory that a transition to a $(\pi,\pi)$ supersolid phase, discussed in the previous sections, happens at this point.

\section{Discussion}

We have studied through linear spin wave theory the various mechanisms which destroy the superfluid phase in the bosonic ring-exchange model Eq.~(\ref{JKham}) at half-filling, motivated as a candidate to realize exotic phases in cold atomic gases in optical lattices.\cite{Buchler}  As is known from previous QMC results,\cite{JK_1st} sufficiently large ring-exchange $K>0$ promotes: first, a quantum phase transition to a valence bond solid state; second, a quantum phase transition to a $(\pi,\pi)$ charge-density wave (CDW) state, at large $K$.  In this paper, we have shown that a moderate value of $K<0$ is also sufficient to destroy the superfluid phase, promoting a $(\pi,\pi)$ ordered phase that is identified as a supersolid, or a phase with coexisting superfluidity, CDW, and valence-bond-like order.

It is interesting to combine the exact results from QMC and our current SW theory results to speculate on the $T=0$ phase diagram for all parameters $K/J$ at half-filling.  
The supersolid phase that we find at $K/J<-2$ must be related to the CDW phase found by exact QMC simulations for $K/J>14$, since in the limit $K/J \rightarrow \infty$, the Hamiltonian Eq.~(\ref{JKham}) for $K<0$ maps exactly onto  the Hamiltonian with $K>0$, via a rotation of spins on one of the sub lattices by $\pi/2$ around the $z$ axis.  For a finite value of $K/J<0$, two possibilities exist for the evolution of the supersolid state to the CDW; the state can remain supersolid for all finite $K/J$, or there could be a quantum phase transition between the supersolid and the $(\pi,\pi)$ CDW solid at some critical $K/J$.  In either case, we have identified in this paper that a supersolid phase is possible at half-filling in the relatively simple Hamiltonian Eq.~(\ref{JKham}) that has already been identified as a candidate for construction in ultracold bosons in optical lattices.\cite{Buchler}

Further, we have studied the finite-temperature phase boundary of the superfluid phase in linear SW theory, by calculating the superfluid density and estimating the Kosterlitz-Thouless transition temperature by forcing it to obey the universal jump condition.  We find that this procedure yields good numerical agreement with exact QMC results for $K=0$.  For $K<0$, the phase boundary monotonically decreases to $T=0$ at $K/J=-2$, indicating promotion of the $(\pi,\pi)$ supersolid phase discussed above.  Increasing $K>0$ from the XY-point, the phase boundary initially increases, reaching a maximum for $K \neq 0$ before monotonically decreasing.  This is in qualitative agreement with the trend identified with QMC simulations,\cite{JKannl} and allows us to attribute at least the initial increase in $T_{KT}$ for small $K/J$ to the physics of non-interacting spin waves.  
It is remarkable that such nontrivial information about this model, namely the non-monotonic behavior of $T_{KT}$ as a function of $K/J$, and the existence of a novel supersolid phase,  
can be contained in such a simple analytical theory.

\section{Acknowledgments}
We are indebted to P. Nikolic, A. Paramekanti, A. Sandvik and D. Scalapino for enlightening discussions.  Support for this work was provided by NSERC of Canada.

\bibliography{JKbiblio}

\begin{thebibliography}{21}
\expandafter\ifx\csname natexlab\endcsname\relax\def\natexlab#1{#1}\fi
\expandafter\ifx\csname bibnamefont\endcsname\relax
  \def\bibnamefont#1{#1}\fi
\expandafter\ifx\csname bibfnamefont\endcsname\relax
  \def\bibfnamefont#1{#1}\fi
\expandafter\ifx\csname citenamefont\endcsname\relax
  \def\citenamefont#1{#1}\fi
\expandafter\ifx\csname url\endcsname\relax
  \def\url#1{\texttt{#1}}\fi
\expandafter\ifx\csname urlprefix\endcsname\relax\def\urlprefix{URL }\fi
\providecommand{\bibinfo}[2]{#2}
\providecommand{\eprint}[2][]{\url{#2}}

\bibitem[{\citenamefont{Paramekanti et~al.}(2002)\citenamefont{Paramekanti,
  Balents, and Fisher}}]{Arun1}
\bibinfo{author}{\bibfnamefont{A.}~\bibnamefont{Paramekanti}},
  \bibinfo{author}{\bibfnamefont{L.}~\bibnamefont{Balents}}, \bibnamefont{and}
  \bibinfo{author}{\bibfnamefont{M.~P.~A.} \bibnamefont{Fisher}},
  \bibinfo{journal}{Phys. Rev. B} \textbf{\bibinfo{volume}{66}},
  \bibinfo{pages}{054526} (\bibinfo{year}{2002}).

\bibitem[{\citenamefont{Sandvik et~al.}(2002)\citenamefont{Sandvik, Daul,
  Singh, and Scalapino}}]{JK_1st}
\bibinfo{author}{\bibfnamefont{A.~W.} \bibnamefont{Sandvik}},
  \bibinfo{author}{\bibfnamefont{S.}~\bibnamefont{Daul}},
  \bibinfo{author}{\bibfnamefont{R.~R.~P.} \bibnamefont{Singh}},
  \bibnamefont{and} \bibinfo{author}{\bibfnamefont{D.~J.}
  \bibnamefont{Scalapino}}, \bibinfo{journal}{Phys. Rev. Lett.}
  \textbf{\bibinfo{volume}{89}}, \bibinfo{pages}{247201}
  (\bibinfo{year}{2002}).

\bibitem[{\citenamefont{Balents and Paramekanti}(2003)}]{Arun2}
\bibinfo{author}{\bibfnamefont{L.}~\bibnamefont{Balents}} \bibnamefont{and}
  \bibinfo{author}{\bibfnamefont{A.}~\bibnamefont{Paramekanti}},
  \bibinfo{journal}{Phys. Rev. B} \textbf{\bibinfo{volume}{67}},
  \bibinfo{pages}{134427} (\bibinfo{year}{2003}).

\bibitem[{\citenamefont{Rousseau et~al.}(2005)\citenamefont{Rousseau,
  Scalettar, and Batrouni}}]{rick1}
\bibinfo{author}{\bibfnamefont{V.~G.} \bibnamefont{Rousseau}},
  \bibinfo{author}{\bibfnamefont{R.~T.} \bibnamefont{Scalettar}},
  \bibnamefont{and} \bibinfo{author}{\bibfnamefont{G.~G.}
  \bibnamefont{Batrouni}}, \bibinfo{journal}{Phys. Rev. B}
  \textbf{\bibinfo{volume}{72}}, \bibinfo{pages}{054524}
  (\bibinfo{year}{2005}).

\bibitem[{\citenamefont{B\"{u}chler et~al.}(2005)\citenamefont{B\"{u}chler,
  Hermele, Huber, Fisher, and Zoller}}]{Buchler}
\bibinfo{author}{\bibfnamefont{H.~P.} \bibnamefont{B\"{u}chler}},
  \bibinfo{author}{\bibfnamefont{M.}~\bibnamefont{Hermele}},
  \bibinfo{author}{\bibfnamefont{S.~D.} \bibnamefont{Huber}},
  \bibinfo{author}{\bibfnamefont{M.~P.~A.} \bibnamefont{Fisher}},
  \bibnamefont{and} \bibinfo{author}{\bibfnamefont{P.}~\bibnamefont{Zoller}},
  \bibinfo{journal}{Phys. Rev. Lett.} \textbf{\bibinfo{volume}{95}},
  \bibinfo{pages}{040402} (\bibinfo{year}{2005}).

\bibitem[{\citenamefont{Motrunich and Fisher}(2007)}]{DBL}
\bibinfo{author}{\bibfnamefont{O.~I.} \bibnamefont{Motrunich}}
  \bibnamefont{and} \bibinfo{author}{\bibfnamefont{M.~P.~A.}
  \bibnamefont{Fisher}}, \bibinfo{journal}{Phys. Rev. B}
  \textbf{\bibinfo{volume}{75}}, \bibinfo{pages}{235116}
  (\bibinfo{year}{2007}).

\bibitem[{\citenamefont{Sheng et~al.}(2008)\citenamefont{Sheng, Motrunich,
  Trebst, Gull, and Fisher}}]{Donna}
\bibinfo{author}{\bibfnamefont{D.~N.} \bibnamefont{Sheng}},
  \bibinfo{author}{\bibfnamefont{O.~I.} \bibnamefont{Motrunich}},
  \bibinfo{author}{\bibfnamefont{S.}~\bibnamefont{Trebst}},
  \bibinfo{author}{\bibfnamefont{E.}~\bibnamefont{Gull}}, \bibnamefont{and}
  \bibinfo{author}{\bibfnamefont{M.~P.~A.} \bibnamefont{Fisher}},
  \bibinfo{journal}{arXiv:0805.0255}  (\bibinfo{year}{2008}).

\bibitem[{\citenamefont{Melko et~al.}(2004{\natexlab{a}})\citenamefont{Melko,
  Sandvik, and Scalapino}}]{JK_h}
\bibinfo{author}{\bibfnamefont{R.~G.} \bibnamefont{Melko}},
  \bibinfo{author}{\bibfnamefont{A.~W.} \bibnamefont{Sandvik}},
  \bibnamefont{and} \bibinfo{author}{\bibfnamefont{D.~J.}
  \bibnamefont{Scalapino}}, \bibinfo{journal}{Phys. Rev. B}
  \textbf{\bibinfo{volume}{69}}, \bibinfo{pages}{100408}
  (\bibinfo{year}{2004}{\natexlab{a}}).

\bibitem[{\citenamefont{Senthil et~al.}(2004)\citenamefont{Senthil, Vishwanath,
  Balents, Sachdev, and Fisher}}]{DQCP1}
\bibinfo{author}{\bibfnamefont{T.}~\bibnamefont{Senthil}},
  \bibinfo{author}{\bibfnamefont{A.}~\bibnamefont{Vishwanath}},
  \bibinfo{author}{\bibfnamefont{L.}~\bibnamefont{Balents}},
  \bibinfo{author}{\bibfnamefont{S.}~\bibnamefont{Sachdev}}, \bibnamefont{and}
  \bibinfo{author}{\bibfnamefont{M.~P.~A.} \bibnamefont{Fisher}},
  \bibinfo{journal}{Science} \textbf{\bibinfo{volume}{303}},
  \bibinfo{pages}{1490} (\bibinfo{year}{2004}).

\bibitem[{\citenamefont{Sandvik and Melko}(2006)}]{JKannl}
\bibinfo{author}{\bibfnamefont{A.~W.} \bibnamefont{Sandvik}} \bibnamefont{and}
  \bibinfo{author}{\bibfnamefont{R.~G.} \bibnamefont{Melko}},
  \bibinfo{journal}{Ann. Phys.} \textbf{\bibinfo{volume}{231}},
  \bibinfo{pages}{1651} (\bibinfo{year}{2006}).

\bibitem[{\citenamefont{Gomez-Santos and Joannopoulos}(1987)}]{GomezJ}
\bibinfo{author}{\bibfnamefont{G.}~\bibnamefont{Gomez-Santos}}
  \bibnamefont{and} \bibinfo{author}{\bibfnamefont{J.~D.}
  \bibnamefont{Joannopoulos}}, \bibinfo{journal}{Phys. Rev. B}
  \textbf{\bibinfo{volume}{36}}, \bibinfo{pages}{8707} (\bibinfo{year}{1987}).

\bibitem[{\citenamefont{Bernardet et~al.}(2002)\citenamefont{Bernardet,
  Batrouni, Meunier, Schmid, Troyer, and Dorneich}}]{TroyerSWT}
\bibinfo{author}{\bibfnamefont{K.}~\bibnamefont{Bernardet}},
  \bibinfo{author}{\bibfnamefont{G.~G.} \bibnamefont{Batrouni}},
  \bibinfo{author}{\bibfnamefont{J.-L.} \bibnamefont{Meunier}},
  \bibinfo{author}{\bibfnamefont{G.}~\bibnamefont{Schmid}},
  \bibinfo{author}{\bibfnamefont{M.}~\bibnamefont{Troyer}}, \bibnamefont{and}
  \bibinfo{author}{\bibfnamefont{A.}~\bibnamefont{Dorneich}},
  \bibinfo{journal}{Phys. Rev. B} \textbf{\bibinfo{volume}{65}},
  \bibinfo{pages}{104519} (\bibinfo{year}{2002}).

\bibitem[{\citenamefont{Sandvik}(1997)}]{AWS_heisenberg}
\bibinfo{author}{\bibfnamefont{A.~W.} \bibnamefont{Sandvik}},
  \bibinfo{journal}{Phys. Rev. B} \textbf{\bibinfo{volume}{56}},
  \bibinfo{pages}{11678} (\bibinfo{year}{1997}).

\bibitem[{\citenamefont{Sandvik and Hamer}(1999)}]{AWShammer}
\bibinfo{author}{\bibfnamefont{A.~W.} \bibnamefont{Sandvik}} \bibnamefont{and}
  \bibinfo{author}{\bibfnamefont{C.~J.} \bibnamefont{Hamer}},
  \bibinfo{journal}{Phys. Rev. B} \textbf{\bibinfo{volume}{60}},
  \bibinfo{pages}{6588} (\bibinfo{year}{1999}).

\bibitem[{\citenamefont{Wessel and Troyer}(2005)}]{ss1}
\bibinfo{author}{\bibfnamefont{S.}~\bibnamefont{Wessel}} \bibnamefont{and}
  \bibinfo{author}{\bibfnamefont{M.}~\bibnamefont{Troyer}},
  \bibinfo{journal}{Phys. Rev. Lett.} \textbf{\bibinfo{volume}{95}},
  \bibinfo{pages}{127205} (\bibinfo{year}{2005}).

\bibitem[{\citenamefont{Heidarian and Damle}(2005)}]{ss2}
\bibinfo{author}{\bibfnamefont{D.}~\bibnamefont{Heidarian}} \bibnamefont{and}
  \bibinfo{author}{\bibfnamefont{K.}~\bibnamefont{Damle}},
  \bibinfo{journal}{Phys. Rev. Lett.} \textbf{\bibinfo{volume}{95}},
  \bibinfo{pages}{127206} (\bibinfo{year}{2005}).

\bibitem[{\citenamefont{Melko et~al.}(2005)\citenamefont{Melko, Paramekanti,
  Burkov, Vishwanath, Sheng, and Balents}}]{ss3}
\bibinfo{author}{\bibfnamefont{R.~G.} \bibnamefont{Melko}},
  \bibinfo{author}{\bibfnamefont{A.}~\bibnamefont{Paramekanti}},
  \bibinfo{author}{\bibfnamefont{A.~A.} \bibnamefont{Burkov}},
  \bibinfo{author}{\bibfnamefont{A.}~\bibnamefont{Vishwanath}},
  \bibinfo{author}{\bibfnamefont{D.~N.} \bibnamefont{Sheng}}, \bibnamefont{and}
  \bibinfo{author}{\bibfnamefont{L.}~\bibnamefont{Balents}},
  \bibinfo{journal}{Phys. Rev. Lett.} \textbf{\bibinfo{volume}{95}},
  \bibinfo{pages}{127207} (\bibinfo{year}{2005}).

\bibitem[{\citenamefont{Melko et~al.}(2006)\citenamefont{Melko, Maestro, and
  Burkov}}]{ss4}
\bibinfo{author}{\bibfnamefont{R.~G.} \bibnamefont{Melko}},
  \bibinfo{author}{\bibfnamefont{A.~D.} \bibnamefont{Maestro}},
  \bibnamefont{and} \bibinfo{author}{\bibfnamefont{A.~A.}
  \bibnamefont{Burkov}}, \bibinfo{journal}{Phys. Rev. B )}
  \textbf{\bibinfo{volume}{74}}, \bibinfo{pages}{214517}
  (\bibinfo{year}{2006}).

\bibitem[{\citenamefont{Melko et~al.}(2004{\natexlab{b}})\citenamefont{Melko,
  Sandvik, and Scalapino}}]{RogXY}
\bibinfo{author}{\bibfnamefont{R.~G.} \bibnamefont{Melko}},
  \bibinfo{author}{\bibfnamefont{A.~W.} \bibnamefont{Sandvik}},
  \bibnamefont{and} \bibinfo{author}{\bibfnamefont{D.~J.}
  \bibnamefont{Scalapino}}, \bibinfo{journal}{Phys. Rev. B}
  \textbf{\bibinfo{volume}{69}}, \bibinfo{pages}{014509}
  (\bibinfo{year}{2004}{\natexlab{b}}).

\bibitem[{\citenamefont{Kosterlitz and Thouless}(1973)}]{KT}
\bibinfo{author}{\bibfnamefont{J.~M.} \bibnamefont{Kosterlitz}}
  \bibnamefont{and} \bibinfo{author}{\bibfnamefont{D.~J.}
  \bibnamefont{Thouless}}, \bibinfo{journal}{J. Phys. C}
  \textbf{\bibinfo{volume}{6}}, \bibinfo{pages}{1181} (\bibinfo{year}{1973}).

\bibitem[{\citenamefont{Nelson and Kosterlitz}(1977)}]{Ujump}
\bibinfo{author}{\bibfnamefont{D.~R.} \bibnamefont{Nelson}} \bibnamefont{and}
  \bibinfo{author}{\bibfnamefont{J.~M.} \bibnamefont{Kosterlitz}},
  \bibinfo{journal}{Phys. Rev. Lett.} \textbf{\bibinfo{volume}{39}},
  \bibinfo{pages}{1201} (\bibinfo{year}{1977}).

\end{thebibliography}

\end{document}